\documentclass{appolb}
\usepackage{epsfig}

\def\np #1#2#3  {{Nucl. Phys.~{\bf #1} (19#3) #2 }}
\def\nc #1#2#3  {{Nuovo. Cim.~{\bf #1} (19#3) #2}}
\def\pl #1#2#3  {{Phys. Lett.~{\bf #1} (19#3) #2}}
\def\pr #1#2#3  {{Phys. Rev.~{\bf #1} (19#3) #2}}
\def\prl #1#2#3  {{Phys. Rev. Lett.~{\bf #1} (19#3) #2}}
\def\prep #1#2#3 {{Phys. Rep.~{\bf #1} (19#3) #2}}
\def\zp #1#2#3  {{Z. Phys.~{\bf #1} (19#3) #2}}
\def\epj #1#2#3  {{Eur. Phys. J.~{\bf #1} (#3) #2}}
\def\rmp #1#2#3  {{Rev. Mod. Phys.~{\bf #1} (19#3) #2}}
\def\JETP #1#2#3 {{Sov.\ Phys.\ JETP~{\bf #1} (19#3) #2}}
\def\sj #1#2#3 {{Sov.\ J.\ Nucl.\ Phys.~{\bf #1} (19#3) #2}}
\def\hepph  #1 {{hep-ph/#1 }}
\def \as {\relax\ifmmode\alpha_s\else{$\alpha_s${ }}\fi}

\def \al #1 {\frac {\as({#1})}{\pi} }
\def \ds #1 {\ooalign{$\hfil/\hfil$\crcr$#1$}}

\def \CO {{\cal O}}

\def \ba{\begin{eqnarray}}
\def \ea{\end{eqnarray}}
\def \be{\begin{equation}}
\def \ee{\end{equation}}

\begin{document}
%\eqsec  % uncomment this line to get equations numbered by (sec.num)
\title{
\begin{flushright}
%\vspace*{-4cm}
\small{
BNL-NT-03/39\\
RBRC-389\\
TTP03-37 \\
SFB/CPP-03-55\\
YITP-SB-03-60}
\end{flushright}
The resummed Higgs boson transverse momentum distribution at the LHC
\thanks{Presented by A.~Kulesza at the XXVII International Conference
  of Theoretical Physics, Ustro\'n, Poland, 15-21 September 15-21 2003.}
}
\author{Anna Kulesza
\address{Institut f\"ur Theoretische Teilchenphysik, Universit\"at
  Karlsruhe, D-76131 Karlsruhe, Germany}
\and
George Sterman
\address{C.N.\ Yang Institute for Theoretical Physics,
Stony Brook University,
Stony Brook, New York 11794 -- 3840, U.S.A.}
\and
{Werner Vogelsang}
\address{RIKEN-BNL Research Center and Nuclear Theory, \\
Brookhaven National Laboratory,
Upton, NY 11973, U.S.A.}
}
\maketitle
\begin{abstract}
We apply QCD resummation techniques to study the transverse momentum
distribution of Higgs bosons produced via gluon-gluon fusion
at the LHC. In particular we focus on the joint resummation
formalism which resums both threshold and transverse momentum
corrections simultaneously. A comparison of results obtained in the
joint and the standard recoil resummation frameworks is presented.
\end{abstract}
%\PACS{PACS numbers come here}
  
\section{Introduction}

The Large Hadron Collider (LHC) is widely expected to be the Higgs
boson discovery machine. In particular, the dominant channel for the 
production of a light Higgs particle at the LHC is gluon fusion
$gg \rightarrow HX$~\cite{LesHouches02}.  The search strategies for the Higgs boson
rely deeply on the knowledge of production characteristics, the
transverse momentum ($Q_T$) of the produced boson 
being one of the most important quantities.
In this talk we describe an application of the joint
resummation formalism~\cite{LSV} to calculate transverse momentum
distribution of Higgs bosons produced through the gluon fusion
mechanism at the LHC.
   
It is a general feature of perturbative calculations in QCD that 
close to a phase space boundary partonic hard-scattering cross sections 
acquire large logarithmic corrections. 
These corrections are  related to soft 
and collinear gluon emission and arise from cancellations between virtual and real
contributions at each order in perturbation theory. 
The threshold and recoil corrections are the two notable examples
often discussed in this context. 
The threshold corrections
of the form $\as^n \ln^{2n-1}(1-z)/(1-z)$ become large when the partonic
c.m. energy approaches the invariant
mass $Q$ of the produced boson, $z=Q^2/\hat s \rightarrow 1$. 
The recoil corrections, in turn, are of the form $\as^n
\ln^{2n-1}(Q^2/Q_T^2)$ and grow large if the transverse momentum
carried by the produced boson is very small, $Q_T \ll Q$. Thus,
sufficiently close to the phase-space boundary, i.e. in the limit of soft
and/or 
collinear radiation, fixed-order perturbation theory is bound to fail.
A proper treatment of higher-order corrections in this
limit requires resummation of logarithmic corrections to all orders.

In the Standard Model the leading $\CO(\as^2)$ process for 
Higgs boson production via gluon-gluon fusion 
proceeds through a heavy quark loop, with the 
top quark loop providing the most significant contribution. In the
limit of large top mass, $m_t\rightarrow \infty$, the Higgs
coupling to gluons through the top loop can be described by the
effective $ggH$ vertex~\cite{NLO}. This simplifying approximation was shown to be
valid up to a few percent accuracy in the case of NLO calculations~\cite{Higgsthreshold}. 
The fixed-order predictions for the total production rate are 
currently known at the NNL (next-to-next-to-leading) order. 
Although not as large as the NLO
corrections~\cite{NLO}, the NNLO corrections were found to be substantial,
increasing the NLO predictions by around 30\%~\cite{NNLO}. Moreover, it was shown that the
prevailing contribution to these corrections corresponds to the soft
and collinear gluon emission~\cite{NNLOsoftcolCdFG,NNLOsoftcolHK},
 thus reinforcing the need for a careful
treatment of logarithmic corrections to all orders.

The resummation techniques are well established both in the threshold~\cite{Sthr,CTthr} 
and in the recoil~\cite{preCSS,CSS} case for the Drell-Yan
production-type processes. The Drell-Yan mechanism and the mechanism
for Higgs boson production through gluon-gluon fusion are
similar. This makes it possible to apply (after implementing necessary changes
accounting for gluons, instead of quarks, in the initial state)
the already developed resummation methods to Higgs boson production.   
The resummed predictions were obtained in~\cite{Higgsthreshold,CdFGN}
for threshold resummation and
in~\cite{HN,Higgsrecoil,BQ,BCdFG} for recoil resummation.  
A joint, simultaneous treatment of the threshold and recoil
corrections was first introduced in~\cite{LSV,LSVprl}. It relies on a
novel refactorization of short-distance and long-distance physics at fixed
transverse momentum and energy~\cite{LSV}. Similarly to standard threshold and 
recoil resummation, exponentiation of logarithmic corrections occurs in the
impact parameter $b$ space~\cite{CSS}, Fourier-conjugate to transverse momentum $Q_T$
space as well as in the Mellin-$N$ moment space~\cite{Sthr,CTthr},
conjugate to $z$ space. The resulting expression respects energy and
transverse momentum conservation. A full phenomenological analysis of
$Z$ boson production at the
Tevatron in the framework of joint resummation can be
found in~\cite{KSV}, whereas Higgs boson production at the LHC was
studied in~\cite{KSV2}.
%In the following we present the
%jointly resummed expression, translated to the Higgs boson production case,
%and discuss numerical results.

%% Consequently, phenomenological evaluation of the
%% joint resummation expressions requires providing prescriptions for inverse
%% transforms from $N$ and $b$ spaces. This also involves  
%% specifying
%% a border between resummed perturbation theory and the nonperturbative
%% regime, by analyzing and parameterizing nonperturbative effects. Moreover, to fully define the expressions a procedure for
%% matching between the fixed-order and the resummed result needs to be specified.
%% These aspects of the joint resummation formalism were discussed in
%% detail in the context of electroweak boson production and can be found
%% elsewhere~\cite{}.  

\section{The jointly resummed cross section}
The general expression for the jointly resummed cross
section~\cite{LSV}, applied to Higgs boson production via 
gluon fusion, reads~\cite{KSV2}
\ba
\label{crsec}
       \frac{d\sigma_{AB}^{\rm res}}{dQ^2\,d^2 \vec Q_T}
       &=&   \pi \tau \sigma^h_{0} \delta(Q^2 -m_h^2)\,
H(\as(Q^2))
\int_{C_N}\, \frac{dN}{2\pi i} \,\tau^{-N}\;    \int \frac{d^2b}{(2\pi 
)^2} \,
e^{i{\vec{Q}_T}\cdot {\vec{b}}}\, \nonumber \\
&\!\!\!\!\!\!\!\!\!\!\!\!\!\!\!\!\!\!\!\!\!\!\!\!\!\!\!\!\!\!\!\!\!\!\!\!\!\!\!\!
\!\!\!\!\!\times& \!\!\!\!\!\!\!\!\!\!\!\!\!\!\!\!\!\!\!\!\!\!\!\!\!
{\cal C}_{g/A}(Q,b,N,\mu,\mu_F )\;
        \exp\left[ \,E_{gg}^{\rm PT}(N,b,Q,\mu)\,\right] \;
        {\cal C}_{g/B}(Q,b,N,\mu,\mu_F) \; ,
\ea
where $\tau =Q^2/S$, $m_h$ is the mass of the Higgs boson, and 
$\pi\tau \sigma^h_{0} \delta (Q^2 -m_h^2)$ 
denotes the lowest order partonic cross section for the process
       $gg\rightarrow HX$ in the limit of large $m_t$, with  
\ba
\sigma^h_0={\sqrt{2} G_F \as^2(m_h)  \over 576 \pi}\;.
\ea

The function $H$ contains the hard virtual part of the perturbative corrections
and up to $\CO(\as)$ is given by~\cite{NLO,dFG}
\be
\label{hfunc}
H(\as)=1+\frac{\as}{2\pi}H^{(1)}=
1+\frac{\as}{2\pi} \left(2 \pi^2+11\right)\; .
\ee
Apart from a lower limit of integration, at the
next-to-leading-logarithm (NLL) accuracy the form of the Sudakov factor 
$E_{gg}^{\rm PT}(N,b,Q,\mu)$ in the jointly resummed
expression~(\ref{crsec}) is the same as for the recoil resummation~\cite{KSV}:
\be
\label{sudakov}
E_{gg}^{\rm PT} (N,b,Q,\mu)=
-\int_{Q^2/\chi^2}^{Q^2} {d k_T^2 \over k_T^2} \;
\left[ A_g(\as(k_T))\,
\ln\left( {Q^2 \over k_T^2} \right) + B_g(\as(k_T))\right]  \; .
\ee

%Dependence on the renormalization scale $\mu$ is implicit in
%Eq.~(\ref{sudakov})
%through the expansion of $\as(k_T)$ in powers of $\as(\mu)$ and
%logarithms of $k_T/\mu$.
The functions $A$ and $B$ are perturbative series in $\as$ and 
their
coefficients can be determined by comparing 
fixed order predictions with an expansion of resummed result~\cite{dFG,KT}.
NLL accuracy requires using $A_g^{(1)},\ B_g^{(1)}$ and 
$A_g^{(2)}$
in Eq.~(\ref{sudakov}):
\ba \label{abcoeffs}
A_g^{(1)} &=&  C_A\; , \qquad\qquad
B_g^{(1)}\;=\;-\frac{1}{6} \left( 11 C_A - 4 T_R N_F 
\right)\;,\nonumber \\
A_g^{(2)} &=& \frac{C_A}{2} \left[
C_A \left( \frac{67}{18}-\frac{\pi^2}{6} \right) -\frac{10}{9}T_R
N_F\right]\; ,
\ea
where $C_A=3$, $C_F=4/3$, $T_R=1/2$, and $N_F$ is the number of
flavors. 
The higher order (needed at NNLL) 
coefficient $B_g^{2}$ is also known~\cite{dFG}
\ba \label{Btwo}
\!\!\!\!\!\!\!\!
B_g^{(2)}&=&
C_A^2\left(-{4 \over 3}+{11 \over 36}\pi^2 -{3 \over2}\zeta_3\right)
+ {1 \over 2}C_F T_R N_F + C_A N_F T_R
\left({2 \over 3} - {\pi^2 \over 9} \right).
\ea

The quantity $\chi(N,b)$ appearing in the lower limit of integration
in~(\ref{sudakov}) is specific to joint resummation. The  LL and NLL
logarithmic terms in the threshold limit, 
$N \rightarrow \infty$ (at fixed $b$), and in the recoil limit 
$b \rightarrow \infty$ (at fixed
$N$) are correctly reproduced with the following choice of the form of 
$\chi$
\be
\label{chinew}
\chi(\bar{N},\bar{b})=\bar{b} + \frac{\bar{N}}{1+\eta\,\bar{b}/
\bar{N}}\; ,
\ee
where $\eta$ is a constant and we define
\ba
\label{nbdefs}
\bar{N} = N{\rm e^{\gamma_E}} \; , \\
\bar{b}\equiv b Q {\rm e^{\gamma_E}}/2 \; ,
\ea
with $\gamma_E$ the Euler constant.

The functions ${\cal C}(Q,b,N,\mu,\mu_F )$ in Eq.~(\ref{crsec})
are given by:
\be
\label{cpdf}
{\mathcal C}_{a/H}(Q,b,N,\mu,\mu_F )
=   \sum_{j,k} C_{a/j}\left(N, \alpha_s(\mu) \right)\,
{\cal E}_{jk} \left(N,Q/\chi,\mu_F\right) \,
               f_{k/H}(N ,\mu_F) \; .
\ee
The product of parton distribution functions $f_{k/H}$ at scale $\mu_F$,
and a matrix ${\cal E}_{jk}$ can be seen as corresponding to parton densities 
evaluated at the scale $Q/\chi$. The evolution from the scale $\mu_F$ 
  to $Q/\chi$ is accurate to NLL in $\chi$ and
represented by the matrix ${\cal E} \left(N,Q/\chi,\mu_F\right)$.
The origin and the structure of the evolution matrix  ${\cal E}$ 
was discussed in detail in Ref.~\cite{KSV,KSV2}.
The coefficients
$C_{a/j}(N,\as)$ have a structure of a perturbative series in
$\as$, and are determined in the same way as for recoil
resummation, i.e. up to $\CO(\as)$,
\ba
\label{ccoeff}
C_{g/g}\left( N,\as \right)&=& 1+ \frac{\as}{2\pi}C_{g/g}^{(1)}
= 1+\frac{\as}{4\pi} \,\pi^2\; ,\\
C_{g/q}\left( N,\as\right) &=& \frac{\as}{2\pi}C_{g/q}^{(1)} =
 \frac{\as}{2\pi} C_F \frac{1}{N+1}\; =\;
C_{g/\bar{q}}\left( N,\as\right)\;  .
\ea

The expression~(\ref{crsec}) is formally accurate to the 
next-to-leading-logarithm (NLL) level. However, due to the large colour charge 
of the incoming gluons, the cross section
exhibits increased sensitivity to the Sudakov logarithms. It is known
for recoil resummation that the NNLL terms have a
significant impact on numerical results. Motivated by this
finding, we decide to include the NNLL terms containing the $B^{(2)}$
coefficient, in the way consistent with the recoil resummation.
 In our formalism we also include the $\CO(\as)$ perturbative
 expansions for the functions $H$ and $C$, which formally give 
NNLL contributions. As discussed in Ref.~\cite{CdFG}, the NNLL resummed
cross sections are resummation scheme dependent, and the choice of the
resummation scheme is reflected in the value of the coefficients
$B^{(2)}, \, H^{(1)}$ and $C^{(1)}$.  We exercise the freedom of the
resummation scheme choice by demanding that the function $H$, calculated 
with $\as$ taken at the scale $Q$, collects the
hard virtual part of the NLO corrections. The Sudakov factor
and the $\cal{C}$ coefficients contain then only soft or collinear
contributions. The values of $B^{(2)}$ and $C^{(1)}$ listed above are
for this particular resummation scheme.

By incorporating full evolution of parton densities the
cross section~(\ref{crsec}) correctly includes also the leading $\as^n
\ln^{2n-1}(\bar N) /N$ collinear non-soft terms to all orders.
At the NLO, this can be seen by expanding jointly resummed cross section
to $\CO(\as)$ (here for illustration purposes only in the $gg$ channel), integrated
over $Q_T$
\begin{eqnarray} \label{sec1}
&&\hat{\sigma}^{gg}= \sigma_g^{(0)} \,\frac{\as}{2\pi}\,\left\{
-4 C_A\ln^2\bar{N}+8 \pi b_0 \ln\bar{N}
+11+3\pi^2\right.
\nonumber \\
&&
-2 \ln\bar{N}
\left.
\left[ \frac{4C_A}{N(N-1)}+\frac{4C_A}{(N+1)(N+2)}-4 C_A S_1(N)
+4 \pi b_0\right] \right\}\; ,
\end{eqnarray}
where $S_1(N)=\sum_{j=1}^N j^{-1}=\psi(N+1)+\gamma_E$, with $\psi$
the digamma function. In the large $N$ limit this gives 
\begin{equation} 
\hat{\sigma}^{gg}= \sigma_g^{(0)} \,\frac{\as}{2\pi}\,\left\{
   4 C_A \ln^2\bar{N}+
4C_A \frac{\ln\bar{N}}{N}+11+3\pi^2 \nonumber \right\}+{\cal O}
\left(\frac{\ln\bar{N}}{N^2} \right)\, ,
\label{larn}
\end{equation}
which can be compared to the the large $N$ limit of the NLO result in
the Mellin space 
\begin{eqnarray} 
  \label{largen2}
\hat{\sigma}^{gg}&=& \sigma_g^{(0)} \,\frac{\as}{2\pi}\,\left\{
   4 C_A \ln^2\bar{N}+
4C_A \frac{\ln\bar{N}}{N}+11+4\pi^2 \right\}+
{\cal O}\left(\frac{\ln\bar{N}}{N^2} \right)\, .
\end{eqnarray}
The agreement between the expanded jointly resummed
expression and the exact NLO result down to the $\CO(1/N)$ is clear.
The mismatch in the constant $\pi^2$ term between~(\ref{larn}) and~(\ref{largen2})  
originates from the value of the $C_{g/g}^{(1)}$ coefficient and is a
NNLL effect. A development of the joint formalism at the NNLL would
eliminate this disagreement as well as provide a way to include other
NNLL coefficients, most notably $B^{(2)}$.
 
%% By incorporating full evolution of parton densities (as opposed to only
%% the leading $N$ part of the anomalous dimension) the
%% cross section~(\ref{crsec}) correctly includes the leading $\as^n
%% \ln^{2n-1}(\bar N) /N$ collinear non-soft terms to all orders.
%% In fact, due to our treatment of evolution, expansion of the resummed cross
%% section~(\ref{crsec}) in the limit $N \rightarrow \infty,\,b=0$ returns all 
%% ${\cal O}(1/N)$ terms in agreement with the fixed-order result.
%% Further comparison can be undertaken in the limit $b \rightarrow
%% \infty,\,N=0$ when our joint resummation turns into standard $Q_T$
%% resummation. Consequently, the NLO transverse momentum distribution is
%% recovered from the ${\cal O}(\as)$ expansion of the jointly resummed cross
%% section exactly in the same way as in the $Q_T$ resummation. Outside of these
%% limits, a numerical 
%% comparison between the fixed-order 
%% and the expanded jointly resummed expression for $d \sigma / d Q_T$ at ${\cal
%% O}(\as)$ shows, as expected, a very good agreement--especially in the small
%% $Q_T$ region.

\vspace*{-2mm}
\subsection{Numerical results and discussion}
\vspace*{-2mm}

In order to have predictive power, the resummed
expression~(\ref{crsec}) needs to be supplemented by a definition 
of the inverse Mellin and Fourier transforms from $N$ and $b$ space.
In the joint approach, the inverse integrals are both treated as contour
integrals in the complex space of $N$ and $b$. 
For the
integrals to be well defined, the contours must not run into the Landau
pole or singularities associated with the form of the function $\chi$.
This procedure provides an
unambigous definition of resummed perturbation theory without an
introduction of additional dimensional scales and implies a functional
form of non-perturbative corrections.
We refer the reader to Ref.~\cite{KSV2} for a detailed discussion 
of the parametrization of the contours.
% and to Ref.~\cite{LSV,KSV}
%for comments on a more general issue 
%of specyfing the border between perturbation theory and the
%non-perturbative regime in the resummation framework.  

The joint resummation formalism with the inverse transforms defined as
contour integrals ensures that predictions can be obtained 
for any non-zero value of $Q_T$. This is not possible in the standard
recoil approach without adding some non-perturbative term of the form
$-gb^2$ to the Sudakov exponent. However, for the purely technical
reasons of numerical stability we also include such a factor  in
our jointly resummed cross section. The value 
of the $g$ parameter, $g=1.67$ GeV$^2$, is adopted from the study in Ref.~\cite{KS}.
However, we checked that the dependence of the results at small $Q_T$ 
on the value of $g$ is negligible, in agreement with what was found 
for the case of pure $b$ space resummation. 
At large $Q_T$, where the $\ln (Q_T^2/Q^2)$ terms taken into account by
resummation lose their importance, it is necessary to match the 
resummed result with a fixed-order result. Here the jointly resummed
result is matched to the $\CO(\as)$ perturbative result~\cite{HN}, in the way
described in~\cite{KSV}.

The numerical results for the Higgs boson transverse momentum
distribution calculated  in the joint resummation
framework were obtained assuming $m_h=125$
GeV, $\mu=\mu_F=Q=m_h$ and using CTEQ5M~\cite{CTEQ5} parton distribution
functions. The parameter $\eta$ in the definition of
$\chi$~(\ref{chinew}) is chosen
to be $\eta=1$. We checked that the numerical dependence of the
predictions on the value of $\eta$ is small~\cite{KSV2}. 

Apart from the joint resummation predictions, in Fig.~1 we also show 
the recoil-only (i.e. $\chi=\bar b$) resummed result. At small to
moderate $Q_T$, the $b$ space
resummed prediction is slightly higher and broader than the one
provided by the joint resummation but the difference is small.
Consequently, we
conclude that the threshold effects are of modest importance at these
values of $Q_T$ and the pure recoil resummation is fully applicable
there.

The $Q_T$-integrated joint cross section, by definition, is expected
to return the threshold resummed result. Although it does so formally 
 up to NLL, numerically the integrated joint distribution returns
a result which is $\sim10$\% lower than the threshold cross section.
We find that this suppresion is caused by subleading
terms included in the joint resummation, more specifically terms
$\propto 1/(N-1)$ in the expansion of the joint expression~(\ref{sec1}).
These terms arise from our treatment of evolution in the coefficients
$\cal{C}$, cf. Eq.~(\ref{cpdf}).
They are important only in the small $N$ limit and therefore not
present in the threshold resummed expression. (However, it is interesting to
observe that the NLO cross section in Mellin space
contains the same subset of terms $\propto 1/(N-1)$ as the expanded 
joint expression taken at $\bar b=0$, plus an additional subset of terms  $\propto
1/(N-1)^2$. The numerical effects of the two subsets cancel almost
entirely, leading to a relatively good approximation of the NLO cross section by
the threshold resummed result.) The small $N$ limit corresponds to 
the limit of small $z=Q^2/\hat s \ll 1$. Given partonic c.m. energies available
at the LHC, the small $z$ terms of the form $\as^n \ln^{(2n-1)}(z)/z$
may indeed play a significant role for light Higgs production. 
These terms can be resummed on their
own~\cite{Haut}, but their full inclusion in the joint formalism alongside
the threshold and recoil corrections requires further work.    
%%%%%%%%%%%%%%%%%%%%%%%%%%%%%%%%%%%%%%%%%%%%%%%%%%%%%%%%%%%
\begin{figure}[h]
\begin{center}
\vspace*{-3mm}
\epsfig{file=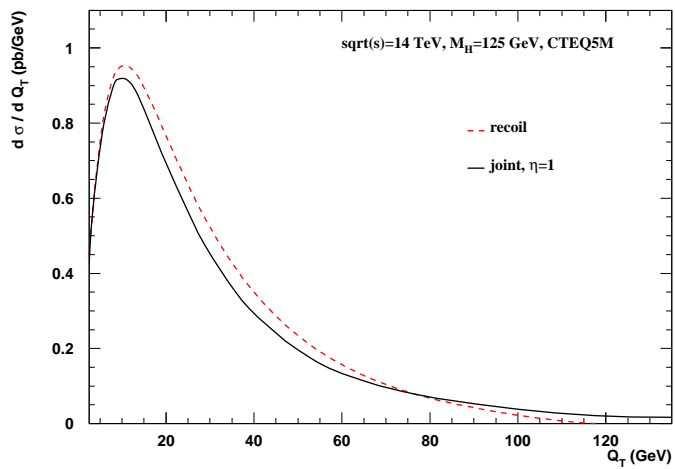,width=10cm}
\end{center}
\vspace*{-5mm}
\caption{Transverse momentum distribution for Higgs production at the LHC in
the framework of joint resummation and of ``pure-$Q_T$'' resummation.}
\label{fig:cdf}
\end{figure}
%%%%%%%%%%%%%%%%%%%%%%%%%%%%%%%%%%%%%%%%%%%%%%%%%%%%%%%%%%%%

\noindent
{\small{\bf Acknowledgements:} A.K. wishes to thank the conference
  organizers for their kind hospitality and a nice atmosphere during
  the meeting. W.V is grateful to RIKEN, Brookhaven National Laboratory 
and the U.S. Department of Energy (contract number DE-AC02-98CH10886) for
providing the facilities essential for the completion of this
  work. The work of G.S. was supported in part by the National Science
  Foundation, grants PHY9722101 and PHY0098527. A.K. acknowledges
  support from Sonderforschungsbereich/Transregio-9 ``Computational
  Particle Physics''.}

\end{document}